\documentclass[pra,aps,twocolumn,superscriptaddress,a4paper]{revtex4}

\usepackage{graphicx} 
\usepackage{epstopdf}

\usepackage{amssymb} 
\usepackage{amsfonts} 
\usepackage{amsmath}
\usepackage{bbold}

\newcommand{\mean}[1]{\langle #1 \rangle}

\newcommand{\w}{\omega}
\newcommand{\W}{\Omega}

\newcommand{\Alpha}{\boldsymbol\alpha}
\newcommand{\Beta}{\boldsymbol\beta}
\newcommand{\Gama}{\boldsymbol\gamma}
\newcommand{\Sigm}{\boldsymbol\sigma}

\begin{document} \title{Dynamics of Gaussian discord between two
oscillators interacting with a common environment}

\author{Jos\'e Nahuel \surname{Freitas}}
\affiliation{Departamento de F\'{\i}sica Juan Jos\'e Giambiagi, FCEyN, UBA,
Pabell\'on 1, Ciudad Universitaria, 1428 Buenos Aires, Argentina}

\author{Juan Pablo \surname{Paz}}
\affiliation{Departamento de F\'{\i}sica Juan Jos\'e Giambiagi, FCEyN, UBA,
Pabell\'on 1, Ciudad Universitaria, 1428 Buenos Aires, Argentina}
\affiliation{Instituto de F\'\i sica de Buenos Aires, IFIBA, UBA CONICET, Pabell\'on 1, Ciudad Universitaria, 1428 Buenos Aires, Argentina}

\begin{abstract} 
We analyze the evolution of the Gaussian discord between two resonant harmonic
oscillators coupled to a common environment. For this, we use the same tools we applied before to fully characterize the evolution of the entanglement in this system (J.P. Paz and A. Roncaglia, Phys. Rev. Lett. 100(2008)). The asymptotic value of Gaussian discord is obtained as a function of parameters characterizing the environment (temperature, couplings, etc) and the initial state of the system (initial squeezing, initial purity, etc). The type of Gaussian measurement optimizing the extraction of information between the oscillators is fully characterized by means of a phase diagram. Such diagram (with phases corresponding to homodyne or heterodyne measurements) has similar topology to the one describing dynamical phases for the entanglement. We present evidence pointing to the fact that Gaussian discord is not always a good approximation of true discord as the asymptotic value of the former is shown to be a non-decreasing function of temperature (in the high temperature regime), reaching an asymptotic value of $\log(2)$ for a pure initial state (and lower values for mixed initial states).  
\end{abstract}

\maketitle

\section{Introduction}

Quantum discord is a measure of quantum correlations that
includes not only those contained in entangled states, but also others that are present in separable states \cite{Ollivier2001}. The fact that separable states can have non-classical correlations, which are quantified by quantum discord, came as a rather bit surprise and has raised considerable attention in recent years. Such attention was partly fueled by recent observations suggesting that discord may also be the computational resource behind the power of the DQC1 model of computation \cite{Knill1998,Datta2008}. Quantum discord has also been related to a variety of problems, such as the complete positivity of quantum
dynamics\cite{Shabani2009} and the state merging protocol\cite{Madhok2011,Cavalcanti2011}, among others. The evolution of quantum discord for quantum open systems has also been studied for systems of 
qubits\cite{Wang2010,Fanchini2010,Werlang2009}, in both the Markovian and
non-Markovian regimes. For continuous variable systems much less is
known. This is because computing the quantum discord is rather hard since it involves solving a optimization problem that is highly nontrivial even for the simplest case of two qubits. However, for a system of two particles (two modes) that are restricted to Gaussian states it is possible to compute an approximation to quantum discord. This is the so-called Gaussian discord, which has been defined and computed recently \cite{Adesso2010,Giorda2010}. 

Here, we fully characterize the behavior of Gaussian discord for a system of two
harmonic oscillators coupled to a common environment.  The environment
is itself a collection of independent harmonic oscillators (linearly coupled
with the system). This same model was used before to study entanglement
dynamics\cite{Paz2009,Paz2008}, where a  complete characterization of the
entanglement in the equilibrium state was given. Our analysis is exact and
includes both non-Markovian and non-perturbative effects. In particular, we analyze here the same scenario
studied before by Maniscalco {\it et al.} in Ref. \cite{Vasile2010}, where the evolution
of Gaussian discord was analyzed using a weak coupling approximation (that, as
we see, prevents one from observing rather interesting behavior such as the
existence of entanglement at long times for initially uncorrelated states). Also, we do not restrict our analysis (as was done in Ref. \cite{Vasile2010}) to high temperatures and include non-Markovian effects well beyond the short time regime. 

The paper is organized as follows: In Sec. \ref{sec:model} we describe the model and the technique, based on the use of the exact master equations.  In
Sec. \ref{sec:discord} we discuss, for completeness, quantum discord and its
Gaussian approximation. In Sec. \ref{sec:results} we present our main
results. We include a full characterization of the behavior of Gaussian discord
in the asymptotic state and study features of the temporal evolution. We
also compare the dynamics of Gaussian discord and entanglement. Finally, Sec. 
\ref{sec:conclusions} contains the conclusions of the work.

\section{The Model} 
\label{sec:model} 

We consider a system of two harmonic oscillators with
coordinates $(x_1,p_1)$ and $(x_2,p_2)$ with the following Hamiltonian (that includes both position and momentum coupling):
\begin{equation}
H_S = \frac{p_1^2}{2m} + \frac{m\w^2x_1^2}{2}
+\frac{p_2^2}{2m} + \frac{m\w^2x_2^2}{2}+ mc_{12} x_1 x_2 +
\frac{\tilde{c}_{12}}{m\w^2}p_1 p_2 
\label{eq:H_sys}
\end{equation}
This system interacts with an environment formed by $N$ independent harmonic oscillators (with a Hamiltonian given by 
$H_{env} = \sum_{n=1}^N \frac{\pi_n^2}{2m_n} + \frac{m_n\w_n^2q_n^2}{2}$. The system-environment interaction is bilineal:
\begin{equation} 
H_{int} = (x_1 + x_2) \sum_{n=1}^N c_n q_n + 
\frac{p_1 + p_2}{m\w} \sum_{n=1}^N \tilde{c}_n \frac{\pi_n}{m\w_n}  
\label{eq:H_int}
\end{equation}  
The total Hamiltonian, $H = H_{sys} + H_{env} + H_{int}$, is quadratic in
position and momentum.  Then, temporal evolution maps initial Gaussian states onto Gaussian
states. Throughout the paper we will consider initial states $\rho_0$ with no correlations between the system and the environment: $\rho_0 = \rho_S \otimes \rho_E$. Moreover, we will assume that the initial state of the environment $\rho_E$ is of thermal equilibrium at certain temperature $T$. The effect of the environment on the system is fully characterized by the initial temperature $T$ and by the spectral density, defined as $J(\w)=\sum_{n=1} ^N \delta(\w-\w_n)
\frac{c_n^2}{2m_n\w_n}$. We consider a family of spectral densities: 
\begin{equation}
J(\w) = \frac{2}{\pi} m\gamma_0 \w \left(\frac{\w}{\Lambda}\right)^{n-1}
\theta(\Lambda -\w) \label{eq:spectral_density} 
\end{equation}
where $\theta(x)$ is the step function. In the continuum limit the environment contains oscillators with frequencies up to the cut-off $\Lambda$. The environment is Ohmic for $n=1$, sub-Ohmic for $n<1$ and super-Ohmic for $n>1$. The strength of the coupling is controlled by the parameter $\gamma_0$. 

As in Ref. \cite{Paz2008}, we consider a family of system-environment interactions with a certain symmetry. This is evident if we rewrite the Hamiltonian in terms of normal modes, whose coordinates are related to those of the original modes as: $x_{\pm} = \frac{1}{\sqrt{2}}(x_1 \pm x_2)$ and $p_{\pm} = \frac{1}{\sqrt{2}} (p_1 \pm p_2)$. Due to the symmetry of the coupling, the interaction with the environment involves only the $x_+$ and $p_+$ operators. In terms of these modes the Hamiltonian of the system is: 
\begin{equation} 
H_S = \frac{p_+^2}{2m_+} +
\frac{m_+\w_+^2x_+^2}{2} +\frac{p_-^2}{2m_-} + \frac{m_-\w_-^2x_-^2}{2}
\label{eq:H_sys_+-} 
\end{equation} 
where $m_{\pm} = m\left(1\pm\frac{\tilde{c}_{12}}{\w^2}\right)^{-1}$  and
$\w_{\pm}^2=\w^2\left(1\pm\frac{c_{12}}{\w^2}\right)\left(1\pm\frac{\tilde{c}_{12}}{\w^2}\right)$
are the masses and frequencies of the new modes, respectively (notice that $+$ and $-$ modes are normal modes since the two oscillators are assumed to be resonant). Two relevant cases can be distinguished within the family of couplings we considered: (1) When the coupling is only trough the position coordinates, i.e., $\tilde{c}_{12} = \tilde{c}_n = 0$. (2) When the coupling is symmetric in position and momentum: $\tilde{c}_n = c_n$.  The
master equation for the state of the system arising for each coupling is well known and is reviewed for completeness in the next subsection. 

\subsection{Position coupling}

In this case, the evolution of the reduced density matrix of the system, $\rho$, is ruled by the exact master equation for Quantum Brownian Motion \cite{HPZ, Paz2009}:
\begin{equation} 
\begin{split} 
\dot{\rho} &= -i[H_R,\rho]-i\gamma(t)[x_+,\{p_+,\rho\}] -\\
 &- D(t)[x_+,[x_+,\rho]] - f(t)[x,[p_+,\rho]] \end{split} 
\label{eq:master_equation_pos} 
\end{equation}
Where $H_R$ is the renormalized Hamiltonian of the system ($ 
H_R = H_S + \frac{m}{2} \delta\W^2(t)x_+^2$). The action of the environment is fully contained in the time dependent coefficients 
$\delta\W^2(t)$, $\gamma(t)$, $D(t)$ and $f(t)$. In turn, these coefficients
can be expressed as functions of the initial temperature $T$ of the environment
and its spectral density. The environment induces a
renormalization on the frequency of the oscillators. The frequency shift
$\delta\W^2(t)$ has a long time limit that depends on the spectral density only
(and depends on the cut-off frequency $\Lambda$). In what follows we choose the
physical parameters in such a way that the long time properties are
cut-off independent. For this, we must choose `bare' parameters $\w^2$ and
$c_{12}$ accordingly (to absorb the cut-off dependency carried by the frequency shift induced by the coupling). For example, in the case of two non-interacting oscillators one must take $c_{12}
= - \lim_{t\rightarrow\infty} \delta\W^2(t)$ so that $\lim_{t\rightarrow
\infty} C_{12}= 0$. For the spectral density (\ref{eq:spectral_density}) one
can show that $\lim_{t\rightarrow\infty} \delta\Omega^2(t) =
-\frac{4}{\pi}\frac{\gamma_0\Lambda}{n}$. The equilibrium value of the
coefficients will be noted simply as $\Omega$, $C_{12}$, $\gamma$, $D$, etc.
The temporal dependence will be written explicitly if needed.

The Gaussian nature of quantum states is preserved by time evolution. This simplifies the description of the evolution since the full state is specified by the first and second moments of the quadrature operators $R=(x_1,p_1,x_2,p_2)$, i.e., by their mean values $\mean{R_i}$ and the covariance matrix $\Sigm_{ij} = \frac{1}{2} \mean{\{R_i,R_j\}} -\mean{R_i}\mean{R_j}$. From the exact master equation (\ref{eq:master_equation_pos}) it is easy to obtain the following evolution equations: 
\begin{equation}
\frac{d}{dt}\mean{p_+} = -m\W_+^2(t)\mean{x_+}-2\gamma(t)\mean{p} \qquad
\frac{d}{dt}\mean{x_+} = \frac{\mean{p_+}}{m} 
\end{equation} 
and: 
\begin{equation} 
\begin{split} 
& \frac{d}{dt}\frac{\mean{p_+^2}}{2m} + \frac{m}{2}\W^2_+(t)\frac{d}{dt}\mean{x_+^2} =
-\frac{2\gamma(t)}{m}\mean{p_+^2}+ \frac{D(t)}{m} \\ 
& \frac{1}{2}\frac{d^2}{dt^2}\mean{x_+^2}+\gamma(t)\frac{d}{dt}\mean{x_+^2}+\W^2_+(t)\mean{x_+^2} =
\frac{\mean{p_+^2}}{m^2}-\frac{f(t)}{m}\\
& \frac{d}{dt}\mean{x_+^2} = \frac{1}{m}\mean{\{x_+,p_+\}} 
\end{split} 
\label{eq:dyn_cov_matrix_pos}
\end{equation} 
Here, $\W_+(t)$ is the renormalized frequency of the $+$ mode.  The evolution
equations for the $-$ mode are simply those of a free harmonic oscillator, and
describe a rotation in the phase space by an angle $\omega_- t$. From the
previous equations it is clear which is the role of the time dependent
coefficients of the master equation: $\gamma(t)$ is a damping coefficient that
induces loss of energy, $D(t)$ is a diffusion coefficient that increases the
dispersion in momentum (and therefore also in position), and $f(t)$ is an 
anomalous diffusion coefficient that affects only the dispersion in position.
This last coefficient reflects the asymmetry of the interaction between system
and environment.

The asymptotic state is attained after a timescale fixed by the long time value of $\gamma(t)$ (of course, this happens after the time dependent coefficients approach constant values, which is the case for typical spectral densities such as the ones we consider here). Thus, from 
(\ref{eq:dyn_cov_matrix_pos}) one can obtain the asymptotic values of the
dispersions $\Delta x_+ = \sqrt{\mean{x_+^2}}$ and $\Delta p_+ =
\sqrt{\mean{p_+^2}}$:
\begin{equation} 
m\W_+\Delta x_+ = \sqrt{\frac{D}{2\gamma}-mf}\qquad \Delta p_+ =
\sqrt{\frac{D}{2\gamma}}
\label{eq:asint_disp_pos}
\end{equation}
Also, the correlations between $x_+$ and $p_+$ vanish in the asymptotic limit:
$\frac{1}{2}\mean{\{x_+,p_+\}}$ = 0. From Eqs. \ref{eq:asint_disp_pos} it
is clear that due to the anomalous diffusion term the equipartition principle
is violated (or, equivalently, $\Delta p_+ \neq m\Omega_+ \Delta x_+$), i.e., there is squeezing in the asymptotic state. This fact have direct consequences in the asymptotic behavior of the correlations between the oscillators.

\subsection{Symmetric coupling} 
When the system-environment interaction is symmetric in position and momentum
($c_n = \tilde{c}_n$), the master equation is \cite{Paz2009}
\begin{equation} 
\begin{split} 
\dot{\rho}(t) =& \frac{1}{i}[H_R,\rho] -i\tilde{\gamma}(t)\left([x_+,\{p_+,\rho\}]-[p_+,\{x_+,\rho\}]\right)-\\
&-\tilde{D}(t)\left([x_+,[x_+,\rho]]+\frac{1}{m_+\w_+^2}[p_+,[p_+,\rho]]\right)
\end{split}
\end{equation} 
Now, the renormalized Hamiltonian is: 
$$H_R = H_S+\delta\tilde{\W}^2(t) \left( \frac{p_+^2}{2m_+\w_+^2}+\frac{m_+x_+^2}{2}\right)$$ 
Again, the time dependent coefficients depend on the spectral density ($\tilde D(t)$ also depends on the temperature). As expected, the
operators $x_+$ and $p_+$ appear in a symmetric way in the master equation (and, as a consequence, there is no anomalous diffusion term). The
asymptotic dispersions are:  
\begin{equation} 
m\W_+\Delta x_+ =\Delta p_+ =\sqrt{\frac{\tilde{D}}{2\tilde{\gamma}}}
\label{eq:asint_disp_sym}
\end{equation}
Where $\tilde{D}$ and $\tilde{\gamma}$ are the asymptotic values of
$\tilde{D}(t)$ and $\tilde{\gamma}(t)$. Also, $\mean{\{x_+,p_+\}}=0$. There is
no squeezing in the asymptotic state of the $+$ mode (i.e., the equipartition
principle is always fulfilled). 

The renormalization in the parameters of the system is as follows:
\begin{equation}
\begin{split}
&m \rightarrow M(t) = m \left(1 + \frac{\delta\W^2(t)}{2\w^2} \right)^{-1}\\
&\w^2 \rightarrow \W^2(t) = \w^2\left(1 + \frac{\delta\W^2(t)}{2\w^2} \right)^2\\
&c_{12} \rightarrow C_{12}(t) =\left(c_{12} + \frac{\delta\W^2(t)}{2}
\right)\left(1 + \frac{\delta\W^2(t)}{2\w^2} \right)\\
&\tilde{c}_{12} \rightarrow \tilde{C}_{12}(t) =\left(\tilde{c}_{12} +
\frac{\delta\W^2(t)}{2} \right)\left(1 + \frac{\delta\W^2(t)}{2\w^2} \right) 
\end{split}
\end{equation}
We considered that the modes $1$ and $2$ do not interact in the long time limit
($C_{12}=\tilde{C}_{12}=0$). With that choice $M=m_-$ and $\Omega = \omega_-$.

\subsection{Asymptotic state} 
\label{sec:asymp_state}
The master equation enables us to obtain not only the asymptotic
covariance matrix of the $+$ mode. In fact, in the same way we can show that all correlations between the $\pm$ modes vanish and that the $-$ mode evolves freely. 
It's covariance matrix, in general, will depend on time. Therefore, the
covariance matrix corresponding to the asymptotic state has the
following form in terms of the $\pm$ coordinates:
\begin{equation}
\Sigm_{\pm}(t)=\begin{pmatrix}(\Delta x_+)^2&&&\\&(\Delta p_+)^2&&\\&&a(t)&c(t)\\&&c(t)&b(t)\end{pmatrix} 
\label{eq:cov_matrix_+-}
\end{equation}
The first $2\times 2$ block on the diagonal is the covariance matrix of the $+$
mode. The second $2\times 2$ block on the diagonal is the covariance matrix of
the $-$ mode, which is obtained as the transformation of an initial diagonal
covariance matrix $\Sigm_-(0) = \left(\begin{smallmatrix}(\Delta
x_-)^2&\\&(\Delta p_-)^2\end{smallmatrix}\right)$ under the rotation in phase
space corresponding to the free evolution: 
\begin{equation} 
\Sigm_-(t) = E(t) \Sigm_-(0) E^T(t) = \begin{pmatrix} a(t)&c(t)\\c(t)&b(t) \end{pmatrix} 
\end{equation} 
where:
\begin{equation}
E(t) = \begin{pmatrix} cos(\w_-t) & \frac{1}{m_-\w_-} sin(\w_-t)\\ -m_-\w_-sin(\w_ t)&cos(\w_- t) \end{pmatrix}
\end{equation}

From this, we can obtain the covariance matrix for the original modes. It has the form $\Sigm_{12} =
\left(\begin{smallmatrix}\Alpha&\Gama\\\Gama^T&\Beta\end{smallmatrix}\right)$
with $\Alpha = \Beta$, reflecting the symmetry of the model. The components of
$\Sigm_{12}$ have simple expressions in terms of $\Delta x_+$, $\Delta p_+$,
$\Delta x_-$, $\Delta p_-$ and $t$, but they are not all necessary. In fact, any measure of correlations must be invariant under
local unitary operations. If these operations are also Gaussian, they
correspond to local symplectic operations at the phase space level. Therefore,
the only quantities one need to know to evaluate measures of correlations like
the mutual information or logarithmic negativity are the local symplectic
invariants of the covariance matrix. These invariants are simply the
determinants of each $2\times 2$ block and the determinant of the entire
covariance matrix. The local symplectic invariants of $\Sigm_{12}$ admit this
compact form: 
\begin{equation}
\begin{split} 
A &= \det(\Alpha) = \frac{1}{4}\left(\phi_+^2+\phi_-^2+2\phi_+\phi_-h(t,r,r_{crit})\right)\\ 
B &= \det(\Beta) = \det(\Alpha) = A\\ 
C &= \det(\Gama) = \frac{1}{4}\left(\phi_+^2+\phi_-^2-2\phi_+\phi_-h(t,r,r_{crit})\right)\\
D &= \det(\Sigm) = \phi_+^2\phi_-^2 
\end{split} 
\label{eq:symplectic_invariants}
\end{equation} 
where $\phi_+ = \Delta x_+ \Delta p_+$ and $\phi_- = \Delta x_- \Delta p_-$ measure the purity of each mode. All the temporal dependence is contained in the function $h(t,r,r_{crit})$: 
\begin{equation}
\begin{split}
h(t,r,r_{crit}) =& \cos^2(\w_- t) \cosh(2(r - r_{crit})) +\\ &\sin^2(\w_- t) \cosh(2(r + r_{crit}))  
\end{split} 
\label{eq:h} 
\end{equation} 
where $r$ and $r_{crit}$ are the squeezing of each mode: 
\begin{equation} 
r_{crit} = \frac{1}{2} \log\left(m_- \w_- \frac{\Delta x_+}{\Delta p_+}\right) \qquad r = \frac{1}{2} \log\left(m_- \w_- \frac{\Delta x_-}{\Delta p_-}\right)
\label{eq:squeeze}
\end{equation} 
From Eq. (\ref{eq:h}) it is evident that if the squeezing of either mode
is zero, then $h$ is constant and there is no temporal dependence in the
symplectic invariants. 

Summarizing, any measure of correlations in the asymptotic state can be
evaluated from the four local invariants A, B, C and D of Eq.
(\ref{eq:symplectic_invariants}). The only memory about the initial state is
contained in the $-$ mode, through $\phi_-$ and $r$. On the other hand, the $+$
mode reaches the thermal equilibrium with the environment, so $\phi_+$ and
$r_{crit}$ depends exclusively on the temperature $T$ and the characteristics
of the environment and its coupling with the system.

\section{Quantum Discord} 
\label{sec:discord} 

Quantum discord arises as the difference between two classically equivalent
expressions for the mutual information of a bipartite system. It can also be
interpreted as the difference between total and classical correlations.
Therefore it is a measure of quantum correlations. Total correlations contained
in a given state $\rho_{AB}$ of a bipartite system are quantified by the mutual
information $I(A:B)$:
\begin{equation} 
I(A:B) = S(\rho_{A}) + S(\rho_{B})-S(\rho_{AB})
\label{eq:mutual_inf} 
\end{equation} 
where $S(\cdot)$ is the Von-Neumman entropy and $\rho_{A(B)}$ is the reduced
state of the subsystem $A$($B$).  Mutual information takes into account all the
correlations, regardless of their nature. Classical correlations are quantified
by the maximum amount of information $J(A:B)$ that can be obtained about a
subsystem, say A, by means of local measurements performed on the other:
\begin{equation} 
\overleftarrow{J}(A:B) = S(\rho_A) - \inf_{\{\Pi_m\}} \left( \sum_m p_m S(\rho_{A|m}) \right)
\end{equation} 
where $\{\Pi_m\}$ is a positive operator-valued measure (POVM) corresponding to a measurement in subsystem B. The
probability of obtaining the result $m$ is $p_m = Tr(\rho_B \Pi_m)$. The state
of $A$ after obtaining this result is $\rho_{A|m} = Tr_B(\rho_{AB}\Pi_m)$.
Then, the sum in the last expression is the average entropy of subsystem $A$
after a measurement described by the POVM $\{\Pi_m\}$ is performed on $B$.
Classically, this sum would coincide with the conditional entropy $S(A|B) =
S(A,B) - S(B)$ and then $I(A:B)$ would be equal to $\overleftarrow{J}(A:B)$ .
The quantum discord $\overleftarrow{D}(A:B)$ is the difference between these
two quantities:
\begin{equation}
\begin{split} 
\overleftarrow{D}(A:B) &= I(A:B) - \overleftarrow{J}(A:B)\\
 &=S(\rho_B) - S(\rho_{AB}) + \inf_{\{\Pi_m\}} \left( \sum_m p_m S(\rho_{A|m}) \right) 
\end{split} 
\label{eq:discord} 
\end{equation}
The main difficulty in calculating the quantum discord of a given state is the
optimization over all the possibles measurements in one subsystem. So far, for
the simplest case of two qubits, an analytical expression for quantum discord
is only available for a reduced family of states. For continuous variable
systems there is only an approximation to the quantum discord, called Gaussian
quantum discord, developed by Adesso and Datta\cite{Adesso2010} and
simultaneously by Giorda and Paris\cite{Giorda2010} in 2010. The main
restriction of this approximation is that the optimization over all the
measurements over one single mode is restricted to the set of Gaussian
measurements, or generalized Gaussian POVM's. For this reason the Gaussian
quantum discord provides only an upper bound for the
unrestricted quantum discord (except for a family of states, for which Gaussian
measurements are known to be optimal). In the following we briefly review the
main points needed to evaluate the Gaussian quantum discord of a two mode
Gaussian state. 

\subsection{Two mode Gaussian states} 
In what follows we consider dimensionless quadrature operators
$R=(x_1,p_1,x_2,p_2)$, i.e., $x_i = \frac{1}{\sqrt{2}}(a_i + a_i^\dagger)$ and
$p_i=\frac{-i}{\sqrt{2}}(a_i + a_i^\dagger)$ where $a_i$ is the annihilation
operator of each mode. 
The Gaussian approximation to the quantum discord applies to two mode Gaussian
states. These are completely described (within local displacements) by the
$4\times 4$ covariance matrix $\Sigm_{12}$ of the quadrature operators :
$$\Sigm_{12} = \begin{pmatrix}\Alpha&\Gama\\\Gama^T&\Beta\end{pmatrix}$$ where
$\Alpha$ and $\Beta$ are the $2\times 2$ covariance matrices for the modes $1$
and $2$, respectively, and $\Gama$ is the matrix containing the correlations
between $(x_1,p_1)$ and $(x_2,p_2)$. As was noted in Sec. 
\ref{sec:asymp_state}, it is necessary only to know the local symplectic
invariants of the covariance matrix. They are $A = \det{\Alpha}$, $B =
\det{\Beta}$, $C = \det{\Gama}$ and $D=\det{\Sigm_{12}}$. Furthermore, by means
of a symplectic operation (generally global), any Gaussian state can be
transformed in another Gaussian state with covariance matrix of the form $\Sigm
= diag(\nu_+,\nu_+,\nu_-,\nu_-)$. The symplectic eigenvalues $\nu_+$ and
$\nu_-$ are invariants under symplectic transformations and can be calculated
as $2\nu_{\pm}^2 = \Delta \pm \sqrt{\Delta^2 - 4 D}$, with $\Delta = A + B +
2C$. A given covariance matrix represents a valid physical state if and only if
$\nu_\pm \geq 1/2$, that is, if the uncertainty principle is satisfied.
Finally, the entropy of a state with symplectic eigenvalues $\{\nu_i\}$ is $S =
\sum_i f(\nu_i)$, with $f(x) = \left(x+1/2\right) \log\left(x+1/2\right) +
\left(x-1/2\right) \log\left(x-1/2\right)$.

\subsection{Gaussian quantum discord} 
The only non trivial point in evaluating the expression (\ref{eq:discord}) for
a two-mode Gaussian state is the optimization of the last term over all the
possible measurement on the mode $2$. The approach of Refs. \cite{Adesso2010,
Giorda2010} consists in restricting the optimization to the set of Gaussian
POVM's. 
The elements of a Gaussian POVM can be expressed as $\Pi(\eta) =
\frac{1}{\pi}W(\eta)\rho_0 W(\eta)^\dagger$ where $\rho_0$ is a valid density
matrix of a single mode Gaussian state, $W(\eta) = e^{\eta a_2 - \eta
a_2^\dagger}$ is the displacement operator and $\eta$ is a complex number
corresponding to each of the possible results. The POVM is completely
determined by $\rho_0$. If the global state is $\rho_{12}$, the probability
density of obtaining the result $\eta$ is $p(\eta) = Tr(\rho_2\Pi(\eta))$ and
the state of the mode $1$ after obtaining that result is $\rho_{1|\eta} =
Tr_2(\rho_{12}(\mathbb{1}\otimes\Pi(\eta)))$. Therefore, the generalization of
the sum in (\ref{eq:discord}) is $\int p(\eta) S(\rho_{1|\eta}) d^2\eta$. It
can be shown that the covariance matrix $\epsilon$ of $\rho_{1|\eta}$ is
independent of $\eta$ and equals $\alpha - \gamma(\beta+\Sigm_0)^{-1}\gamma^T$,
where $\Sigm_0$ is the covariance matrix of the state $\rho_0$ that defines the
POVM. Then $S(\rho_{1|\eta}) = f(\det(\epsilon))$ and the last integral is
trivial. Since $f(\cdot)$ is a growing function, the final step is to minimize
$E = \det(\epsilon)$ over all the covariance matrices $\Sigm_0$. The result is
as follows:
\begin{equation}
\overleftarrow{D}(\rho_{12}) = f(\sqrt{B}) - f(\nu_+) - f(\nu_-) +
f(\sqrt{E_{min}})
\label{eq:gaussian_discord}
\end{equation}
Where $E_{min}$ is given by:
\begin{equation}
\begin{split}
&E_{min} = \min_{\Sigm_0} det(\epsilon) =\\
&= \begin{cases} 
\frac{2C^2+(1/4-B)(A-4D)+2|C|\sqrt{C^2+(1/4-B)(A-4D)}}{4(1/4-B)^2}& \text{if } g
\leq 0, \\
&\\
\frac{AB-C^2 + D - \sqrt{C^4+(-AB+D)^2 - 2C^2(AB+D)}}{2B} & \text{if } g > 0 
\end{cases} 
\label{eq:E_min}
\end{split}
\end{equation}
The quantity $g$ that discriminates the two cases in the last Eq. is:
\begin{equation}
g = (D-AB)^2 - (1/4+B)C^2(A+4D)
\label{eq:discriminant}
\end{equation}
The case $g>0$ corresponds to states for which the Gaussian discord is optimized
by \emph{homodyne} measurements. These measurements are projections onto pure
states of infinite squeezing. For example, the measurement of position or
momentum is homodyne. In the other hand, $g\leq 0$ corresponds to
\emph{heterodyne} measurements. These are projections onto squeezed thermal
states.

\section{Results} \label{sec:results}

This section is divided into three parts. First we review previous
results on entanglement dynamics. In the second part we analyze the Gaussian
discord of the asymptotic state. Finally we discuss the time evolution of discord for intermediate times. 

\subsection{Asymptotic entanglement}

\begin{figure} 
\begin{center}
\includegraphics[scale=.67]{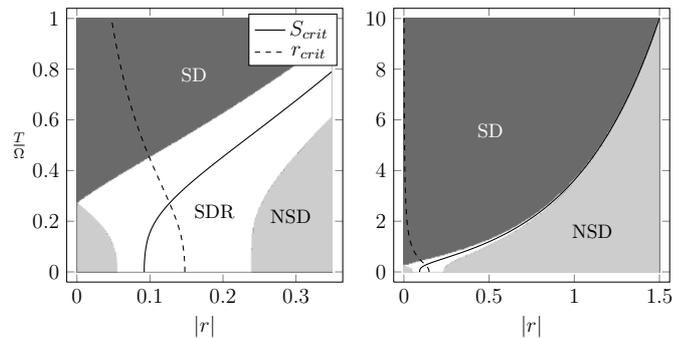} 
\end{center}
\caption{\label{fig:phases_ent_pos}. Phase diagram for the asymptotic
entanglement. Ohmic environment and position coupling. $\Lambda = 20$, $\gamma = 0.1$, $m=1$, $\Omega=1$, $C_{12}=0$.} 
\end{figure}

Entanglement dynamics was investigated in Refs. 
\cite{Paz2008,Paz2009}. A good entanglement measure is given by the logarithmic
negativity, defined as $E_N = \max\{0,-\log(2\tilde{\nu}_-)\}$, where
$\tilde{\nu}_-$ is the smallest symplectic eigenvalue of the covariance matrix
that is obtained after partial transposition. This eigenvalue can be
calculated as $2\tilde{\nu}_-=\Delta - \sqrt{\Delta^2-4\det(\Sigm_{12})}$,
$\Delta = A + B -2C$. As shown in Refs. \cite{Paz2008, Paz2009}, the logarithmic negativity of the asymptotic state is:
\begin{equation}
E_N(t) = \max\{0,E(t)\} 
\label{eq:log_negativity}
\end{equation}
where $E(t)$ is:
\begin{equation}
E(t) = \tilde{E}_N + \Delta E_N \; G(t)
\end{equation}
The function $G(t)$ is an oscillatory function with period $\pi/\omega_-$. The mean value $\tilde{E}_N$ and the amplitude $\Delta E_N$ of $E(t)$ are:
\begin{equation}
\begin{split}
\tilde{E}_N &= \max\{|r|,|r_{crit}|\} - S_{crit}\\
\Delta E_N &= \min\{|r|,|r_{crit}|\}
\end{split}
\end{equation}
The squeezing factors $r$ and $r_{crit}$ are defined in Eq. (\ref{eq:squeeze}),
and $S_{crit}$ is given by $S_{crit}=\frac{1}{2}\log(4\Delta x_+ \Delta p_+
\Delta x_- \Delta p_-)$. From this, it is clear that we can identify three different `phases' for the asymptotic entanglement. First, if $E(t)\le 0$ the asymptotic state is not entangled. This phase is called SD (Sudden Death).
On the other hand, if $E(t)>0$ the asymptotic state is entangled. This
phase is called NSD (No Sudden Death). Finally, there is an intermediate
situation in which the function $E(t)$ alternates between positives and
negatives values. In this case the asymptotic entanglement suffers successive events of `sudden death' and `sudden revivals', and so this phase is called SDR (Sudden Death and Revivals). These three phases are illustrated in Fig.
\ref{fig:phases_ent_pos} for the case of position coupling with an Ohmic
environment. For any initial squeezing
$r$, there exists a temperature above which the asymptotic state enters the SD
phase, i.e., it is separable. This critical temperature grows exponentially with
the initial squeezing $r$, which is a resource to create entanglement. For low temperatures a surprising result emerges (as it is non-perturbative, it is missed by perturbative treatments \cite{Vasile2010}): There is an NSD island which is due to the fact that the asymptotic state is squeezed. For symmetric coupling the phase diagram is much simpler: Only two phases (SD and NSD) exist. In what follows we will see that the dynamics of discord shows many differences and some similarities with the above results. 

\subsection{Asymptotic Gaussian discord}

\subsubsection{Symmetric coupling}
\begin{figure} 
\begin{center}
\includegraphics[scale=1]{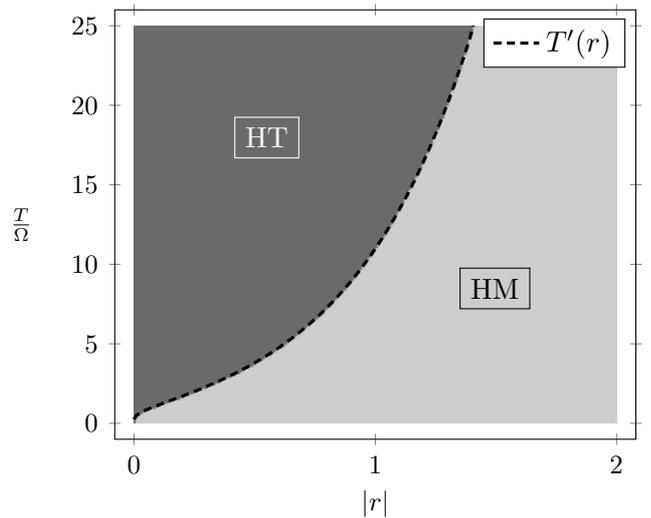} 
\end{center}
\caption{\label{fig:phases_sym}Type of measurement that minimizes the Gaussian
discord of the asymptotic state. HM: homodyne measurements. HT: heterodyne
measurements. Ohmic environment and symmetric coupling. $\Lambda = 20$, $\gamma
= 0.1$, $M=1$, $\Omega=1$, $C_{12}=\tilde{C}_{12}=0$.} 
\end{figure}

The Gaussian discord of the asymptotic state can be evaluated using Eqs. (\ref{eq:gaussian_discord}), (\ref{eq:E_min}), and
(\ref{eq:discriminant}) together with the expressions for the symplectic invariants in Eq. (\ref{eq:symplectic_invariants}). For symmetric coupling the $+$ mode is not squeezed ($r_{crit}=0$) and all the symplectic invariants become time independent (i.e., $h(t,r,r_{crit})=\cosh(2r)$). For simplicity, we
consider the non interacting case ($C_{12} =
\tilde{C}_{12} = 0$) and restrict ourselves to initially pure states for the moment ($\phi_-=1/2$). There are two interesting aspects of quantum discord in the asymptotic regime. First, we analyze what measurement optimizes the extraction of information between the two components of the system (this is needed to obtain the actual value of the discord). After that, we look at the asymptotic value of Gaussian discord as a function of squeezing and temperature. 

Figure \ref{fig:phases_sym} shows the type of measurement, homodyne or
heterodyne, that minimizes the Gaussian discord of the asymptotic state. It is
clear that this phase diagram has the same topology than the one characterizing
the dynamics of the entanglement in the asymptotic regime. For any given
temperature, there is a degree of squeezing after which the optimal measurement
is always homodyne (HM). Contrarily, heterodyne (HT) measurements are optimal
for temperatures above a certain curve $T'(r)$. This is a natural result, in fact in the limit of infinite squeezing there is an observable associated with the relative motion between the two oscillators that is perfectly localized. For
example, if the observable $q_- = \frac{1}{\sqrt{2}}(x_1-x_2)$ is perfectly
localized and the position of mode $2$ is measured, then the state of mode 1 collapses to an eigenstate of $x_1$. This is a pure state with zero
entropy. Therefore, the (homodyne) measurement of the position in mode 2
minimizes the discord since the minimum possible value of the sum of Eq.
(\ref{eq:discord}) is achieved. It is worth pointing out that because the
asymptotic state for our system is symmetric, the discord doesn't depend on the
choice of the subsystem that is measured, as is generally the case.  A simple
expression for the curve $T'(r)$ can be obtained from Eqs. (\ref{eq:discriminant}) and (\ref{eq:symplectic_invariants}). With $\phi_-=1/2$
and $r_{crit}=0$, the discriminant $g$ of eq. (\ref{eq:discriminant}) is a
polynomial of degree 8 in the variables
$\phi_+=\frac{1}{2}\coth(\frac{\Omega}{2T})$ and $h=\cosh(2r)$. The curve
$T'(r)$ correspond to points for which $g=0$. This happens if and only if
$h=h_{crit}(\phi_+)$, with:
\begin{equation}
h_{crit}(\phi_+) = \frac{-1}{3} \phi_+ + \frac{1}{6} \sqrt{16 \phi_+^2 + 56 + \phi_+^{-2}} -\frac{1}{12\phi_+} 
\end{equation}
The function $T'(r)$ is defined implicitly by $\cosh(2r)=h_{crit}(\phi_+(T'))$.
For high temperature or squeezing $\phi_+ \simeq \frac{T}{\Omega}$,
$h_{crit} \simeq \frac{\phi_+}{3}$ and $\cosh(2r) \simeq \frac{e^{2r}}{2}$. Then
$T'(r) \simeq \frac{3\Omega}{2} e^{2r}$.

\begin{figure} 
\begin{center} 
\includegraphics[scale=1]{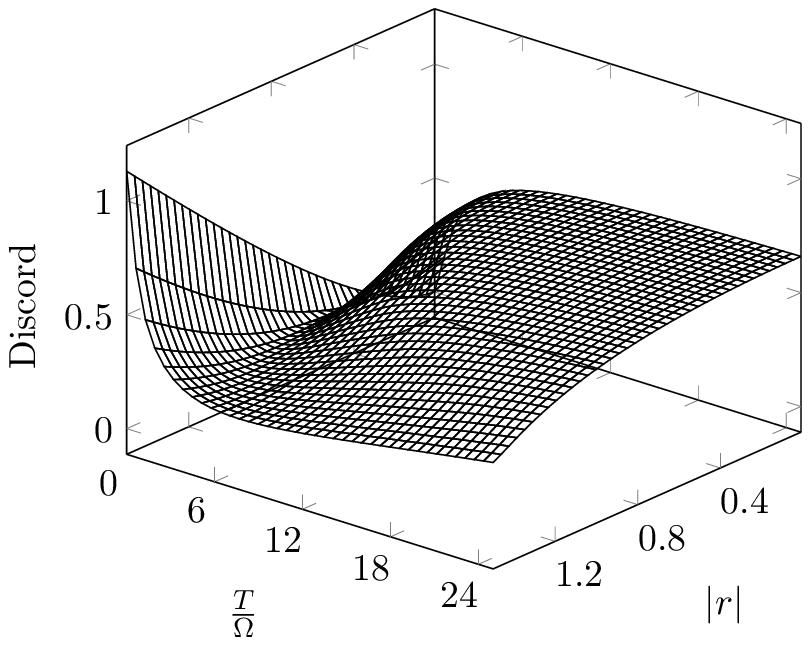}
\end{center} 
\caption{\label{fig:disc_sym}Gaussian discord for the equilibrium state. Ohmic environment and symmetric coupling. $\Lambda = 20$, $\gamma = 0.1$, $M=1$, $\Omega=1$, $C_{12}=\tilde{C}_{12}=0$.} 
\end{figure}

\begin{figure} 
\begin{center}
\includegraphics[scale=1]{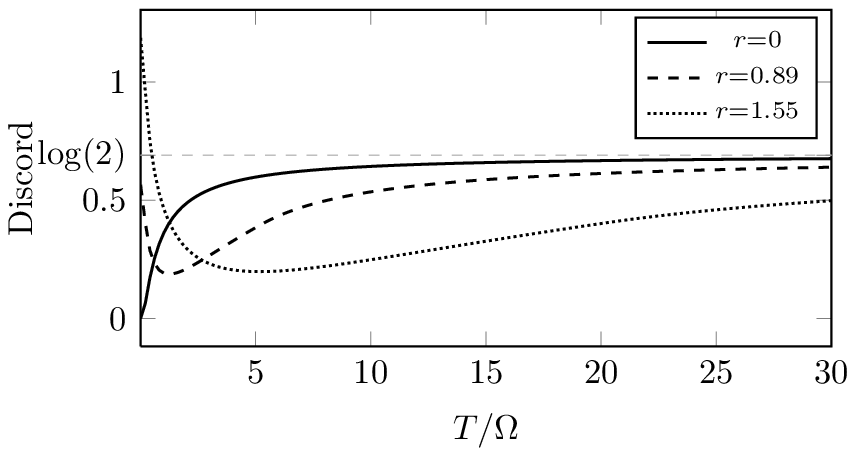} 
\end{center}
\caption{\label{fig:disc_vs_T_sym}Gaussian discord for the equilibrium state. Ohmic environment and symmetric coupling. $\Lambda = 20$, $\gamma = 0.1$, $M=1$, $\Omega=1$, $C_{12}=\tilde{C}_{12}=0$.} 

\end{figure}

It is simple to show that discord can never disappear in a finite time (i.e., there is no sudden death of discord). Figure \ref{fig:disc_sym} shows the asymptotic value of the Gaussian discord as a function of the initial squeezing $r$ and the temperature $T$. As expected, for low temperatures Gaussian discord grows with the squeezing. On the other hand the temperature dependence is rather surprising: There is a temperature after that Gaussian discord grows with temperature approaching a saturation value which can be shown to be equal to $\log(2)$. This is not expected as discord is a measure of quantum correlations, which decrease with temperature. This result can be analytically found for high temperatures: Thus, in such a case the symplectic invariants are $A=B\simeq C\simeq \frac{\phi_+^2}{4}$, and $D=\phi_+^2\phi_-^2$. Then, $E_{min} \simeq 4 (\phi_- + 1/4)^2$ and the Gaussian discord is: 
\begin{equation}
D_{sup}=f\left(\phi_+/2\right)-f\left(\phi_+\right)-f\left(\phi_-\right)+f\left(2\left(\phi_-+1/4\right)\right)
\label{eq:disc_sup}
\end{equation}
As $f\left(\phi_+/2\right)-f\left(\phi_+\right)\rightarrow  -\log(2)$, for initially pure states (for which $\phi_-=1/2$), we have $D_{sup}=\log(2)$, as
Fig. \ref{fig:disc_vs_T_sym} indicates. This high temperature result for Gaussian discord is independent of the asymptotic squeezing of the $+$ mode. Therefore, it is also valid for the case of position coupling that will be discussed below. For high temperatures the optimal Gaussian measurement is always heterodyne and in this case the non-orthogonality of Gaussian states seem to pose a constraint on the maximal information that one can gather about one system by performing Gaussian detection on the other. It seems that this is the cause of the failure of Gaussian discord as a good approximation to true discord (which, as a measure of quantum correlations, must decrease with growing temperature in the high temperature regime).

\subsubsection{Position coupling}
\label{sec:results_equil_state_pos}
\begin{figure} 
\begin{center}
\includegraphics[scale=.67]{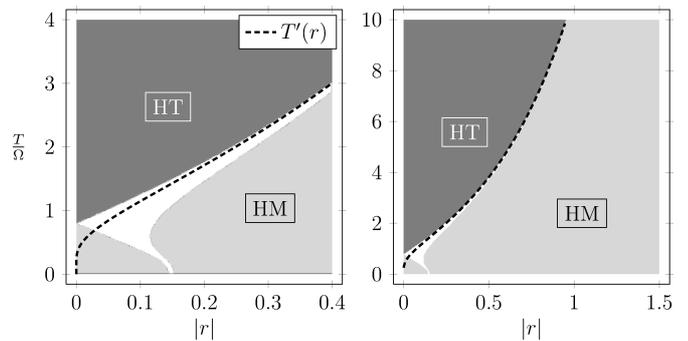} 
\end{center}
\caption{\label{fig:phases_pos} Type of measurement that minimizes the Gaussian
discord for the asymptotic state. Position coupling and Ohmic environment. $\Lambda = 20$, $\gamma = 0.1$, $m=1$, $\Omega=1$, $C_{12}=0$.}
\end{figure}

If the system-environment interaction is through position the asymptotic state
is squeezed (the squeezing is measured by $r_{crit}$). As a consequence, the
local symplectic invariants of the asymptotic state become time dependent (they
oscillate periodically with frequency $\omega_-$, as seen in Eqs. (\ref{eq:symplectic_invariants}) and (\ref{eq:h})). The type of measurement that optimizes the extraction of information between the two subsystems changes because of this fact. This is seen in Figure
\ref{fig:phases_pos}. As opposed to the case discussed above, the phase diagram characterizing the optimal measurement shows three distinct phases. Again, large squeezing and high temperatures are, respectively, associated with HM and HT measurements. However, the fact that the asymptotic state is squeezed due to the coupling induces a low temperature phase where the optimal measurement is homodyne. The existence of an explicit time dependence in the asymptotic state reflects on the fact that the HM and HT phases are separated by a new phase where the optimal measurement depends on time (for some times the optimal measurement is homodyne while for other times such measurement is heterodyne). This is rather similar to what happens with the phase diagram characterizing the evolution of the entanglement in the long time limit. The curve
$T'(r)$ that separates the two phases for the case of symmetric coupling is
plotted for reference. The new time dependent phase disappears for high temperatures (this is due to the fact that for the Ohmic environment we are considering $r_{crit}\rightarrow 0$ for increasing temperature). 
\begin{figure} 
\begin{center}
\includegraphics[scale=1]{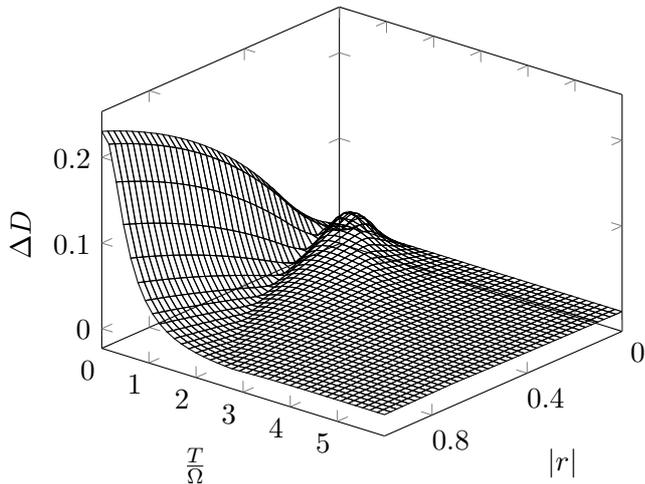} 
\end{center}
\caption{\label{fig:disc_ampl_pos}Peak-to-peak amplitude for the oscillations of
the Gaussian discord in the equilibrium state. Position coupling and
Ohmic environment. $\Lambda = 20$, $\gamma = 0.1$, $m=1$, $\Omega=1$, $C_{12}=0$.} 
\end{figure}

The asymptotic value of discord in this case is also time dependent: Gaussian discord oscillates with $\omega_-$ frequency. The mean value of the
oscillations behaves essentially like the Gaussian discord for the symmetric
coupling case (Fig. \ref{fig:disc_sym}). Figure \ref{fig:disc_ampl_pos} shows
the peak-to-peak amplitude $\Delta D$ of the oscillations of the Gaussian
discord in the equilibrium state. Again, because $r_{crit}$ goes to zero
rapidly for increasing temperature, the amplitude $\Delta D$ of the oscillations
becomes negligible for temperatures that are a few times larger than $\Omega$.

\subsubsection{Mixed initial states}

All the previous results were obtained for pure initial states, i.e., $\phi_-=1/2$. Now we explore how the results change if this condition is relaxed. For
simplicity, only symmetric coupling and Ohmic environments are considered. 
\begin{figure} 
\begin{center}
\includegraphics[scale=.67]{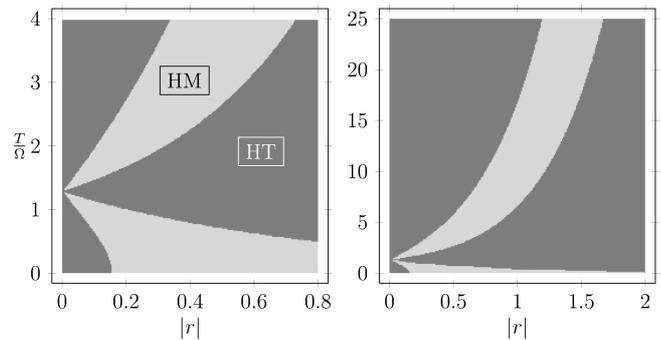} 
\end{center}
\caption{\label{fig:phases_sym_term}Type of measurement that minimizes the
Gaussian discord for a mixed initial state with $\phi_-=3/2$. Symmetric
coupling and Ohmic environment. $\Lambda = 20$, $\gamma = 0.1$, $M=1$, $\Omega=1$, $C_{12}=\tilde{C}_{12}=0$.}
\end{figure}
\begin{figure} 
\begin{center}
\includegraphics[scale=1]{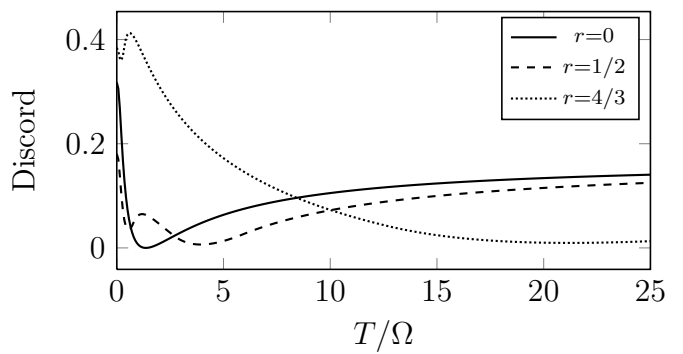} 
\end{center}
\caption{\label{fig:disc_vs_T_term} Gaussian discord of the asymptotic state
versus the temperature of the environment. Mixed initial state with
$\phi_-=3/2$. Symmetric coupling and Ohmic environment. $\Lambda = 20$, $\gamma = 0.1$, $M=1$, $\Omega=1$, $C_{12}=\tilde{C}_{12}=0$.} 
\end{figure}

\begin{figure} 
\begin{center}
\includegraphics[scale=1]{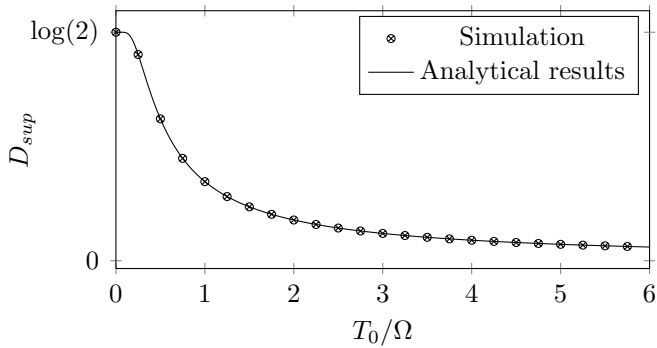} 
\end{center}
\caption{\label{fig:disc_vs_T0} High temperature value of the Gaussian discord
of the asymptotic state versus the initial temperature of the system. The
analytical results (Eq. (\ref{eq:disc_sup})) are compared to numerical simulations.} 
\end{figure}

The nature of the measurement minimizing the Gaussian discord is shown in Fig. \ref{fig:phases_sym_term} for a mixed initial state with $\phi_-=3/2$. There are obvious differences between this and the corresponding figure for a pure initial state (see Figure \ref{fig:phases_sym}). Some of the results are also counter intuitive: for example, in the limit of large squeezing the optimal measurement turns out to be heterodyne. The asymptotic value of discord decreases when increasing the initial mixedness. This is shown in Figure
\ref{fig:disc_vs_T_term} where the asymptotic value of the Gaussian discord is
shown as a function of the temperature for different values of the initial
squeezing (for $\phi_-=3/2$). The asymptotic value of Gaussian discord in the
high temperature limit can be analytically estimated using equation
\ref{eq:disc_sup}. Thus, the asymptotic value of Gaussian discord is examined
as a function of the initial mixedness of the state (which is estimated by the
effective temperature derived from $\phi_-$, $T_0$, which is such that $\phi_- = \frac{1}{2}\coth(\frac{\Omega}{2T_0})$). This is shown in Figure
\ref{fig:disc_vs_T0}. 

\subsection{Dynamics of Gaussian Discord}

Temporal evolution of the Gaussian discord can be analyzed by a simple exact numerical solution of the dynamical equations of the covariance matrix. Results are presented below for two sets of initial conditions corresponding to separable and entangled states for the original modes. 

\subsubsection{Symmetric coupling.}

\begin{figure} 
\begin{center}
\includegraphics[scale=1]{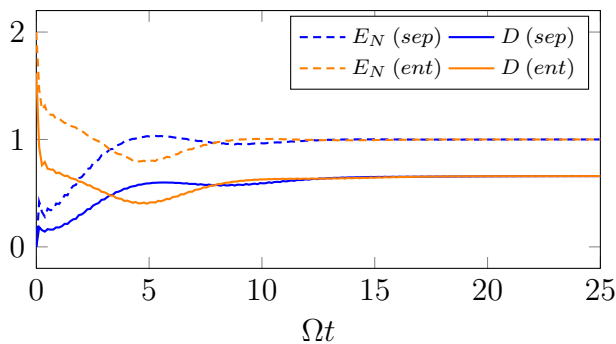} 
\end{center}
\caption{\label{fig:evolution_sym_1} (Color online) Gaussian discord and entanglement
evolution for an entangled and a separable initial state, both with initial
squeezing $r=1$. Symmetric coupling and Ohmic environment at $T=0$. $\Lambda = 20$, $\gamma = 0.1$, $M=1$, $\Omega=1$, $C_{12}=\tilde{C}_{12}=0$.}
\end{figure}

\begin{figure} 
\begin{center}
\includegraphics[scale=1]{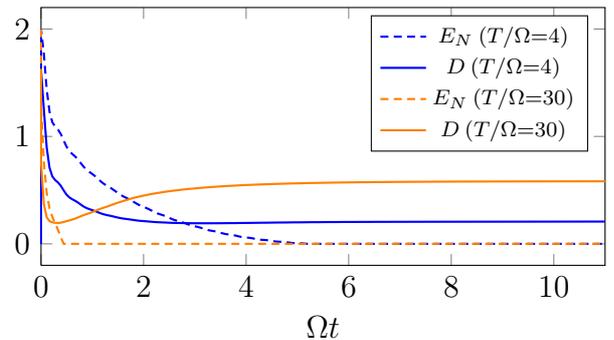} 
\end{center}
\caption{\label{fig:evolution_sym_3} (Color online) Gaussian discord and entanglement
evolution for an entangled initial state with $r=1$. Symmetric coupling and
Ohmic environment at $T=4\Omega$ and $T=30\Omega$. $\Lambda = 20$, $\gamma = 0.1$, $M=1$, $\Omega=1$, $C_{12}=\tilde{C}_{12}=0$.}
\end{figure}

From our previous discussion it is clear that discord in the asymptotic state depends only on the initial squeezing $|r|$ of the $-$ mode. This is
illustrated in Fig. \ref{fig:evolution_sym_1}, where it is shown that Gaussian discord behaves in the same way for both initially separable and entangled states with the same degree of squeezing ($r=1$ in this case). Three temporal scales are 
relevant for the evolution of quantum correlations. For example, for an initially entangled state decoherence leads to a rapid decay of both entanglement and Gaussian discord. This takes place in a timescale that is much shorter than the
dynamical time of the system $\tau = 2\pi/\Omega$. After this, quantum
correlations oscillate with a decreasing amplitude. The timescale
characterizing the oscillations is fixed by $\Omega$ while the amplitude decay
is governed by $\gamma$. For the case of a $T=0$ environment, the evolution of Gaussian discord and entanglement is essentially the same. This is not true for higher temperatures, as Fig. \ref{fig:evolution_sym_3} shows. As seen in the Figure, even though there is sudden death of entanglement, the Gaussian discord reaches an non-zero asymptotic value. For sub-Ohmic environments the results are similar, and
the only difference is that the equilibrium state is reached faster. This is because for a sub-Ohmic environment the coupling with the low frequency oscillators in the environment is stronger.

\subsubsection{Position coupling.}

\begin{figure} 
\begin{center}
\includegraphics[scale=1]{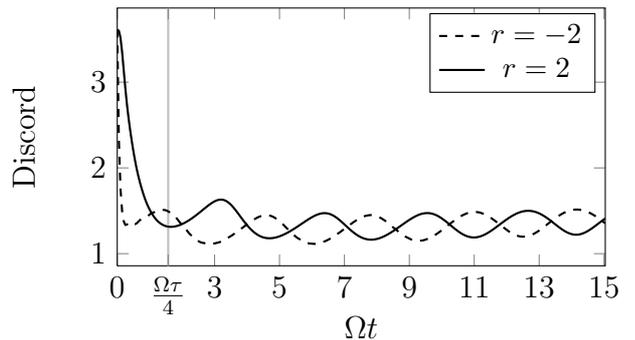} 
\end{center}
\caption{\label{fig:evolution_pos_1} Gaussian discord for two initial states
with positive and negative squeezing. Position coupling and Ohmic environment at $T=0$. $\Lambda = 20$, $\gamma = 0.1$, $m=1$, $\Omega=1$, $C_{12}=0$.}
\end{figure}

As discussed before, when system and environment interact through
position, there are oscillations in the asymptotic
state. The evolution of the Gaussian discord for entangled initial states is shown  in Fig. \ref{fig:evolution_pos_1}. As before, it is seen that the mean value and amplitude of the oscillations depends only on the initial squeezing. For this case, the asymmetry between position and momentum can be seen by looking at the evolution of initial states that are squeezed either along position or along momentum. In fact, as shown in the Figure, initial states with momentum squeezing ($r<0$) are more sensitive to the environmental interaction during the initial time. Such states decohere more rapidly than states that are squeezed along position ($r>0$). For those states to decohere, dynamics has to take place (during which position and momentum exchange roles). This is clearly seen in Fig. \ref{fig:evolution_pos_1}. The position-momentum asymmetry is also seen in the $\pi/2$ phase difference in the long time oscillations of both entanglement and discord. 

\begin{figure} 
\begin{center}
\includegraphics[scale=1]{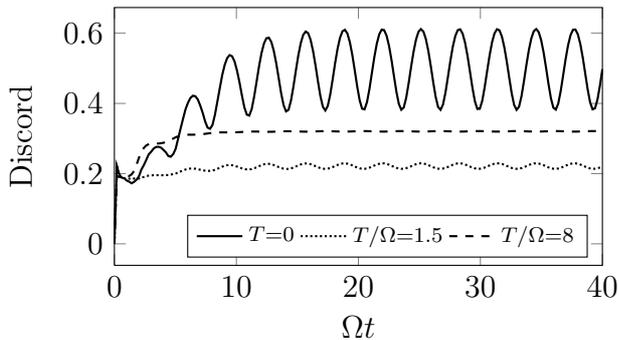} 
\end{center}
\caption{\label{fig:evolution_pos_2} Gaussian discord evolution for an initial
separable state with $r=1$ and different temperatures for the environment. Position coupling and Ohmic environment. $\Lambda = 20$, $\gamma = 0.1$, $m=1$, $\Omega=1$, $C_{12}=0$.}
\end{figure}

\begin{figure} 
\begin{center}
\includegraphics[scale=1]{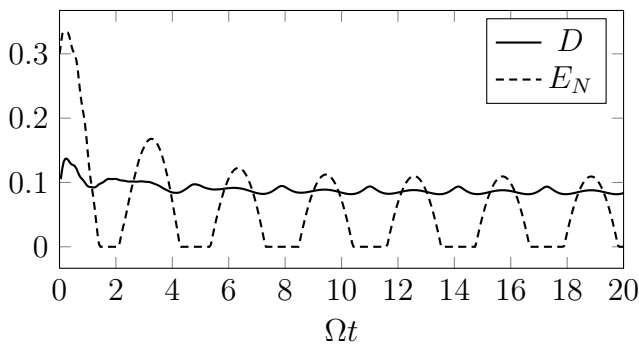} 
\end{center}
\caption{\label{fig:evolution_pos_3} Gaussian discord and entanglement
evolution for an initial state and temperature corresponding to the SDR phase
($T=0.35\Omega$ and $r=0.15$). Position coupling and Ohmic environment. $\Lambda = 20$, $\gamma = 0.1$, $m=1$, $\Omega=1$, $C_{12}=0$.}
\end{figure}

Figure \ref{fig:evolution_pos_2} shows the evolution of Gaussian discord for 
initial separable states and different temperatures. As the
temperature increases the amplitude decreases, in
accordance with the results of Sec. \ref{sec:results_equil_state_pos}.
Finally, Fig. \ref{fig:evolution_pos_3} compares the evolution of
entanglement and Gaussian discord for an initial state and temperature
corresponding to the SDR phase (see Fig. \ref{fig:phases_ent_pos}). It is
interesting to note that the Gaussian discord attains local maximum values when
the entanglement disappears.

\section{Conclusions} \label{sec:conclusions}

We analyzed the dynamics of Gaussian discord of a bipartite system of two
harmonic oscillators that interact with a common bosonic environment. For a
coupling that is bilinear (in position and momentum of both the system and the
environment) the Gaussian measurement optimizing the extraction of information
between the two subsystems can be obtained and used to compute the asymptotic
discord. In this way, the evolution of Gaussian discord can be fully
characterized in the long time limit. Finally, by means of a numerical exact
solution of dynamical equations we analyzed the time dependence of discord for
intermediate times (before the asymptotic state is reached). In this way, we
confirmed analytical evaluation mentioned above. The results have some
surprising features. Previous works on entanglement dynamics showed that such
quantum correlations can be created by the interaction with the common
reservoir (in fact, non-entangled initial states exhibit long time resilient
entanglement). However, this is the case only for temperatures that are low
enough (for any value of the initial squeezing there is a critical temperature
above which the final state is separable). This is not what happens with the type of quantum correlations characterized by discord. Gaussian discord can also be created by the interaction with a common reservoir. However, once it is created by the interaction with the environment, it never disappears. As shown in our work, the asymptotic value of Gaussian discord may be small but never vanishes (as expected). The fact that the asymptotic value for Gaussian discord does not decrease (but saturates) when the temperature grows is far from intuitive. Most likely, this is an indication of the failure of the Gaussian approximation for discord in this case. True quantum discord, as a measure of quantum correlations must decrease with increasing temperature. 

It is interesting to notice that our treatment enables us to obtain many exact results exploring extensively the non-perturbative, non-Ohmic and non-Markovian regimes. In fact, analytic results for entanglement in the asymptotic regime can be obtained using only the long time values of the coefficients that appear in the exact master equation. Such values are known for a variety of physically relevant cases (see Ref. \cite{Paz2009} for a collection of such results). Numerical evaluation is needed only to take care of intermediate time regimes (until the dependence on the initial state is washed out by the evolution). This treatment is in contrast with previous works, such as the one reported in Ref. \cite{Vasile2010} that make extensive use of numerical solution. In particular, using our treatment it is rather simple to relax other previous assumptions (such as the restriction to low coupling or the limitation to short time in the non-Markovian case) such as the ones used in Ref. \cite{Vasile2010}.

We gratefully acknowledge support of grants from CONICET, UBACyT and ANPCyT. 

\bibliographystyle{unsrt} \bibliography{discordPRA}

\end{document}